\documentclass[letterpaper,twocolumn,10pt]{article}
\usepackage{usenix2019_v3}

\usepackage{subfigure}
\usepackage{tikz}
\usepackage{amsmath}
\usepackage{makecell}
\usepackage{amssymb, graphicx, xcolor}
\usepackage{booktabs, colortbl, multirow}
\usepackage{algorithm}
\usepackage{algpseudocode}
\usepackage{listings}
\usepackage{xspace}
\usepackage{enumitem}
\usepackage{xurl}

\definecolor{none}{RGB}{240, 240, 240} 
\definecolor{both}{RGB}{198, 210, 239} 
\definecolor{porcupine}{RGB}{253, 174, 107} 
\definecolor{pine}{RGB}{49, 130, 189} 

\usetikzlibrary{arrows.meta, positioning}
\usetikzlibrary{matrix,decorations.pathreplacing}
\usetikzlibrary{calc}
\usetikzlibrary{fit, backgrounds,shapes}

\tikzset{
  txt/.style={
    draw=none,
    align=center,
    font=\footnotesize
  },
  box/.style={
    draw,
    rounded corners,
    minimum width=1.2cm,
    minimum height=1cm,
    align=center,
    font=\footnotesize
  },
  trace/.style={
    box,
    minimum width=3.3cm,
    minimum height=2cm,
    fill=white
  },
  system/.style={
    box,
    fill=green!5
  },
  arrow/.style={
    ->,
    thick
  }
}

\newcommand{\eg}{{\em e.g.,}\xspace}

\newcommand{\etc}{{\em etc.}\xspace}
\newcommand{\ie}{{\em i.e.,}\xspace}

\newcommand{\code}[1]{{\tt #1}}

\newenvironment{enum}
  {\begin{enumerate}[itemsep=1pt, topsep=2pt,leftmargin=*]}
  {\end{enumerate}}

\newcommand{\op}[1]{{op$_#1$}}
\newcommand{\wop}[1]{{w$_#1$}}
\newcommand{\rop}[1]{{r$_#1$}}
\newcommand{\hint}[1]{\textcolor{pine}{\em #1}}
\newcommand{\name}{\texttt{Pine}\xspace}
\newcommand{\porcupine}{\texttt{Porcupine}\xspace}

\newcommand{\circled}[1]{\textcircled{\scriptsize #1}}

\newcommand{\history}[1]{\mathcal{#1}}
\newcommand{\rb}[1]{\prec_{\history{#1}}^{rb}}

\newcommand{\hard}{\prec_{h}}
\newcommand{\soft}{\prec_{s}}

\newcommand{\faster}{370x\xspace}
\newcommand{\bugs}{6\xspace} 
\newcommand{\uniqbugs}{5\xspace} 
\newcommand{\hintloc}{70\xspace}

\newcommand{\numsut}{8\xspace}

\newcommand{\prut}{Raft, HoliPaxos, EPaxos, LazyLog, Gryff, and Zookeeper\xspace}

\begin{document}

\date{}

\title{\Large \bf Generalizing and accelerating consistency checking for
non-transactional distributed storage systems}

\author{
{\rm Kotikala Raghav} \qquad
{\rm Aman Hassan} \qquad
{\rm Brian Sajeev Kattikat} \\
{\rm Patel Jay} \qquad
{\rm RSRS Santhosh} \qquad
{\rm Abhilash Jindal} \\[0.2cm]
{\rm Indian Institute of Technology Delhi} \\
{\rm \texttt{ajindal@cse.iitd.ac.in}}
}

\maketitle

\begin{abstract}
Linearizability checkers check if an operation history, observed by
concurrent clients, is linearizable. They are used in testing distributed
storage systems, and use the classic Wing-Gong (WG) linearizability
checking algorithm. 

In this paper, we generalize the WG algorithm to make
linearizability checkers more versatile: we can check other non-transactional
consistency guarantees, like ordered sequential consistency provided by
Zookeeper. Equipped with this generalization, we can also check for {\em
system-specific consistency guarantees} that introduce additional ordering
constraints over operations in a history, as per the system's specification.

Our experiments with \numsut distributed storage systems show that checking for 
system-specific consistency guarantees is easy to realize, reduces false 
negatives in testing, helps debug consistency violations, can be upto \faster
faster, and can scale to more concurrent clients within the same checking time
budget. We report \bugs new consistency violation bugs, out of which \uniqbugs
could not be found with existing consistency checkers.
\end{abstract}

\section{Introduction}
\label{sec:intro}

Consistency guarantees constrain the set of behaviors that can be observed by
clients, concurrently interacting with a storage system. Distributed storage
systems have been often found to violate their consistency guarantees by
projects like Jepsen, even though these systems are usually formally specified
and verified. These bugs happen because either (a) system's specification and
implementation diverge~\cite{pcomposition}, or (b) the verification process did not consider
certain types of faults~\cite{empirical_study_eurosys17}, or (c) the model checker checked a smaller
model; the bug reveals itself in longer runs with a particular sequence of
interleavings~\cite{epaxos_tla}.

To find bugs, Jepsen
does fault injection testing. It spawns distributed storage system replicas
and clients, and introduces ``nemeses" that inject server failures,
packet delays, network partitions, clock skew, and other
real-world disruptions, to stress test the system.  The requests and responses
observed by the clients, called an operation history, is passed to {\em consistency
checkers} that check for consistency violations.

This paper fills two gaps in the current non-transactional consistency checkers. 

Firstly, distributed storage systems provide various consistency guarantees to
their users~\cite{zookeeper, rss, spanner, pnuts, tao, vukolic, cosmosdb}.  However, the
current non-transactional consistency checkers, Porcupine~\cite{porcupine} and
Knossos~\cite{knossos}, only check for linearizability.  For example, a Jepsen
test used Knossos to check the operation histories produced by
Zookeeper~\cite{jepsen-zk-blog}. However, since Zookeeper does not provide
linearizability, the analysis was not considered meaningful~\cite{jepsen-zk-gh}.

Secondly, these checkers often struggle to scale, since determining whether a
history is linearizable is known to be NP-complete~\cite{npComp}. For example,
Porcupine issue\#6~\cite{porcupineIssue} reports that a history with just 252 operations causes
the checker to hang. The history has a read following many concurrent writes. In
such scenarios, the checker may have to consider many permutations of the
preceding writes to determine which linearization is consistent with the read's
return value. 

This paper improves upon both these gaps. 

To realize a more versatile consistency checker, we generalize the Wing-Gong
linearizability checking algorithm. This generalization enables us to check a
non-transactional consistency guarantees, stronger than sequential
consistency~\cite{vukolic}, 
such as ordered sequential consistency (Zookeeper~\cite{zk_osc}), regular
sequential consistency (Gryff-RSC~\cite{sosp21_rsc}), and bounded staleness
consistency (CosmosDB~\cite{cosmosdb,vukolic}).

Further, this improved versatility of our checker enables us to check {\em
system-specific consistency guarantees} for systems providing strong
consistency\footnote{Following the definition from~\cite{principles_of_ec}, we
call consistency guarantees stronger than sequential consistency {\em strong}.}.
For instance, say we are testing HoliPaxos~\cite{holipaxos}.  Instead of just
checking for linearizability, we can check for ``HoliPaxos-linearizability"
which poses {\em additional ordering constraints} on operations, based on
HoliPaxos' specification, \eg an operation committed with a lower instance
number should execute earlier. 

To check system-specific consistency guarantees, we add {\em
system-specific ordering hints}, \eg instance number for HoliPaxos, within
responses. We only consider returning ordering hint that is eagerly
available while responding, \ie providing ordering hints shall not introduce
additional network communication and thus additional failure patterns. We
empirically demonstrate that we can add such ordering hints for a variety of 
storage systems in less than \hintloc lines of code.

Because of these additional system-specific ordering constraints over
operations, checking for system-specific consistency guarantee is {\em stronger}
than checking for the basic consistency guarantee. We report 3 bug case studies 
where the operation histories pass linearizability checking, but fail
system-specific consistency checking.
We also found that the additional ordering constraints can be helpful in finding
the root cause of the problem. 

Finally, the additional system-specific ordering constraints also help in accelerating
the checker, as the checker now has to try fewer operation permutations. In our
experiments, we found system-specific consistency checking can be upto \faster
faster, and can support more concurrent clients within the same checking time
budget, compared to the basic consistency checking.

This paper makes the following contributions:
\begin{itemize}[nosep,leftmargin=*]
	\item We generalize the WG linearizability checking algorithm to check for a
	broader set of non-transactional consistency guarantees that are stronger than
	sequential consistency. We show how this generalized algorithm can be used for
	various consistency guarantees, such as the ordered sequential consistency
	provided by Zookeeper.
	\item We propose checking system-specific consistency guarantees as it
	provides stronger checks, helps in finding root-causing issues, runs faster, and scales 
	to more concurrent clients.  We
	identify ordering hints and articulate the system-specific consistency
	guarantees for distributed storage system protocols like \prut.
	\item We implement the ideas presented in this paper in a research prototype
	\name. We inject faults into \numsut non-transactional distributed storage systems
	via Jepsen and check operation histories with \name. We find that checking for
	system-specific consistency guarantees makes \name run upto \faster faster and
	scales better with increasing number of clients,
	compared to checking the basic consistency guarantees. 
	\item In our experiments, \name found \bugs new consistency violation bugs. Out
	of them, \uniqbugs bugs could not be found by existing checkers.
\end{itemize}

\name will be made open source.


\section{Background}
\label{sec:back}

For completeness, this section defines linearizability and the interface of a
linearizability checker. 
Readers familiar with them may skip to Section~\ref{sec:motiv}.



\subsection{Linearizability}

Linearizability is the strongest consistency guarantee, that is typically supported by
non-transactional distributed storage systems~\cite{herlihy}.  Consistency
guarantees are defined as {\em constraints} on {\em operation histories}.
An operation history (or a trace) consists of
several operations, many of which may be concurrent to one another.  Each
operation in the history has an invocation time, a return time, the operation's
type and parameters, and the operation's
response\footnote{\label{ft:incomplete}For brevity, we do not discuss
incomplete histories, where responses for some operations are
missing. We refer interested readers to~\cite{principles_of_ec}.}. 

\tikzset{
    default tikz/.style={>=stealth, baseline=(current bounding box.center), scale=0.75},
}

\newcommand{\interval}[7][]{
    
    \draw[thick] (#2,#3) -- (#2+#4,#3);
    \draw[thick] (#2,#3-0.1) -- (#2,#3+0.1);
    \draw[thick] (#2+#4,#3-0.1) -- (#2+#4,#3+0.1);
    
    \node[above] at (#2+#4/2,#3) {#5};
    
    \if B#7
        \node[below] at (#2+#4*0.17647,#3-0.1) {#6};
    \else
        \if L#7
            \node[left] at (#2,#3) {#6};
        \else
            \node[right] at (#2+#4,#3) {#6};
        \fi
    \fi
    
    \if\relax\detokenize{#1}\relax
    \else
        \draw[thick] (#2+#4*#1,#3) node {$\bullet$};
    \fi
}

\begin{figure*}[h]
  	\centering
	\begin{tabular}{@{}c@{\hspace{0.8em}}c@{\hspace{0.8em}}c@{}}
		\subfigure[\label{fig:lin_eg}]{
			\begin{tikzpicture}[default tikz]
				\interval[0.5]{0}{5}{2}{$put(0)$}{\op{1}}{L}	
				\interval[0.2]{1}{4}{2}{$get() = 0$}{\op{3}}{L}	
				\interval[0.5]{0.5}{3}{4}{$put(1)$}{\op{2}}{L}		
				\interval[0.8]{3.5}{5}{1.5}{$get() = 1$}{\op{4}}{L}	
			\end{tikzpicture}
		}
		&
		\begin{tabular}{c}
			\subfigure[\label{fig:seq_eg}]{
				\begin{tikzpicture}[default tikz]
					\interval{0}{2}{1.7}{$put(0)$}{\op{1}}{B}		
					\interval{1.9}{2}{1.7}{$get() = 0$}{\op{3}}{B}	
					\interval{3.8}{2}{1.7}{$put(1)$}{\op{2}}{B}		
					\interval{5.7}{2}{1.7}{$get() = 1$}{\op{4}}{B}	
				\end{tikzpicture}
			}
			\\[1.5em]
			\subfigure[\label{fig:invalid_eg}]{
				\begin{tikzpicture}[default tikz]
					\interval{0}{2}{1.7}{$put(0)$}{\op{1}}{B}		
					\interval{1.9}{2}{1.7}{$put(1)$}{\op{2}}{B}		
					\interval{3.8}{2}{1.7}{$get() = 0$}{\op{3}}{B}	
					\interval{5.7}{2}{1.7}{$get() = 1$}{\op{4}}{B}	
				\end{tikzpicture}
			}
		\end{tabular}
		&
		\subfigure[\label{fig:nonlin_eg}]{
			\begin{tikzpicture}[default tikz]
				\interval{0.5}{4}{1.5}{$put(0)$}{\op{1}}{L}		
				\interval{3}{4}{2}{$get() = 0$}{\op{3}}{R}		
				\interval{0}{3}{2.5}{$put(1)$}{\op{2}}{L}		
				\interval{3}{2}{2.5}{$get() = 1$}{\op{4}}{L}	
			\end{tikzpicture}
		}
	\end{tabular}
	\vspace{-10pt}
	\caption{Example histories on a register.
	(a) A linearizable history.
	(b) A valid sequential history.
	(c) An invalid sequential history. 
	(d) A non-linearizable history.}
	\vspace{-15pt}
\end{figure*}
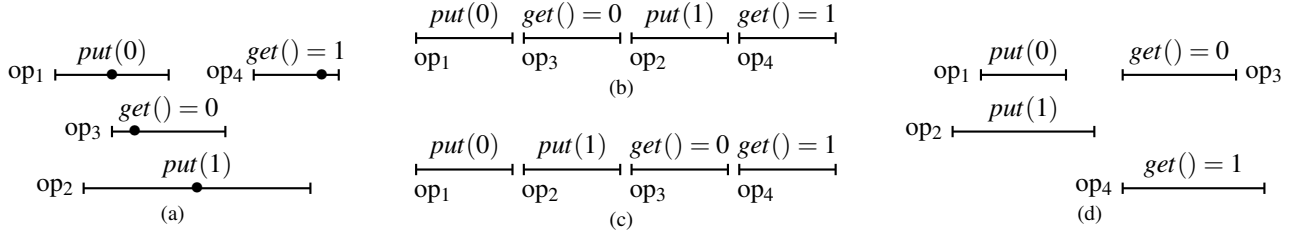

For example, Figure~\ref{fig:lin_eg} shows a history $\history{H}$ with 4
operations on a concurrent register. \op{1} puts 0. \op{3} gets 0. Time advances
from left to right; each operation is shown as an interval with an invocation
time and a response time. Each horizontal level represents the operations of a single client.

The operation history $\history{H}$ determines a {\em return-before order}
between its operations, which we denote as $\rb{H}$.  For example in
Figure~\ref{fig:lin_eg}, \op{1} $\rb{H}$ \op{4} since \op{1} {\em returns
before} \op{4} starts. Whereas, neither \op{2} $\rb{H}$ \op{3} nor \op{3}
$\rb{H}$ \op{2} holds, because \op{2} and \op{3} are concurrent. The
return-before order is a {\em total order} for {\em sequential} histories.
Figure~\ref{fig:seq_eg} shows a sequential history.

Burckhardt~\cite{principles_of_ec} calls a history $\history{H}$ linearizable if
it satisfies three primitive guarantees:
\begin{enum}
  \item \texttt{SingleOrder}: The history $\history{H}$ should be equivalent to
  a sequential history $\history{S}$ \ie both $\history{H}$ and $\history{S}$
  should have the same set of operations, with identical parameters, operation
  type, and return values\textsuperscript{\ref{ft:incomplete}}.
  \item \texttt{RVal}: The equivalent sequential history $\history{S}$ must be
  {\em valid}, \ie read operations return latest written value.
  \item \texttt{RealTime}: If \op{1} $\rb{H}$ \op{2}, then \op{1} $\rb{S}$
  \op{2}.
\end{enum}


For example, the operation history $\history{H}$ in Figure~\ref{fig:lin_eg} is
linearizable because:
\begin{enum}
  \item \texttt{SingleOrder}: The history $\history{H}$ is equivalent to a
  sequential history $\history{S}$, shown in Figure~\ref{fig:seq_eg}. 
  Both histories have the same 4 operations, and each operation preserves its
  response.
  \item \texttt{RVal}: The sequential history $\history{S}$ in
  Figure~\ref{fig:seq_eg} is valid as each read returns the latest written
  value. 
  \item \texttt{RealTime}: The return-before order of operations in
  $\history{S}$ (\op{1} $\rb{S}$ \op{3} $\rb{S}$ \op{2} $\rb{S}$ \op{4}) is
  consistent with the return-before order of operations in $\history{H}$
  (\op{1},\op{3} $\rb{H}$ \op{4}).
\end{enum}

Whereas, the operation history in Figure~\ref{fig:nonlin_eg} is {\em not} linearizable. It
satisfies \texttt{SingleOrder} with the sequential histories in both
Figure~\ref{fig:seq_eg} and Figure~\ref{fig:invalid_eg} as all of them have the
same set of operations. However, when mapping it to Figure~\ref{fig:seq_eg}, it
fails \texttt{RealTime}: \op{2} $\rb{H}$ \op{3}, but \op{3} $\rb{S}$ \op{2}.
When mapping it to Figure~\ref{fig:invalid_eg}, it fails \texttt{RVal}: \op{3}
reads $0$ which is not the latest written value of $1$ by \op{2}. 

\subsection{Linearizability checking}
A {\em linearizability checker}~\cite{porcupine, knossos, alexhorn, wg, lowe}
takes as input an operation history and a {\em sequential specification} of the object,
and determines whether the history is linearizable.  The checker must decide
whether there exists a {\em valid total order} of operations that is {\em
consistent} with the history's return-before partial order.  The sequential
specification defines what it means for a total order of operations to be {\em
valid}. 

For example, Listing \ref{lst:porcupine_reg} shows a sequential specification of
a register initialized to 0. This specification is taken from Porcupine, a
state-of-the-art linearizability checker, used by etcd, TiDB, Amazon MemoryDB,
and others~\cite{porcupine}. The specification allows two operations: a
\code{put} and a \code{get} (lines 1--4). The \code{Step} function determines
whether a particular operation (\code{input}), is allowed to get the specified
\code{output} in its current \code{state} (lines 12--23). For the register, the
\code{Step} function specifies that the \code{put} operation always succeeds and
sets the value of the register (lines 16--17)  while the \code{get} operation is
allowed only if it outputs the current state of the register (lines 19--21).

\begin{lstlisting}[language=Go, float=t, belowcaptionskip=-15pt,
	caption={Sequential specification of a register in the Porcupine
	linearizability checker, written in Go.},
	label=lst:porcupine_reg]
type registerInput struct {
  op    bool // false = put, true = get
  value int
}
registerModel := porcupine.Model{
  Init: func() interface{} {
    return 0
  },
  // Step: takes a state, input, and output, 
  // returns whether it was a legal operation, 
  //  along with a new state
  Step: func(state, input, output interface{}) 
	            (bool, interface{}) {
    regInput := input.(registerInput)
    if regInput.op == false {
      return true, regInput.value 
      // always ok to execute a put
    } else {
      readCorrectValue := output.(int) == state
      return readCorrectValue, state 
      // state is unchanged
    }
  },
}
\end{lstlisting}

Porcupine and others~\cite{alexhorn,knossos} rely on optimized versions of Wing-Gong (WG)
linearizability checking algorithm~\cite{wg}. Briefly, the WG algorithm inputs the
operation history as a doubly linked list. Each operation in the history has a
call and a return. All calls and returns are sorted in the doubly linked list by
real-time. The algorithm tries to serialize all the operations into a stack. When
the algorithm is unable to serialize an operation by the time the operation
returns, the algorithm backtracks by popping some operations from the stack, and
trying a different sequence of operations. We refer the readers to
~\cite{wg, lowe, pcomposition} for more details.

\section{Limitations of existing checkers}
\label{sec:motiv}

\subsection{Versatility}
Distributed storage systems provide a wide variety of consistency guarantees.
For example, K-bounded staleness consistency, such as provided by
CosmosDB~\cite{cosmos-consistency-levels}, relaxes just the \texttt{RVal} of
linearizability to \texttt{K-BoundedStalenessRVal}: a read can return any of the
previous $K$ writes. With this change, the previously invalid sequential history
in Figure~\ref{fig:invalid_eg} is now {\em valid} as \op{3} is now allowed to do
a (bounded) stale read of 0. Hence, the non-linearizable history in
Figure~\ref{fig:nonlin_eg} is allowed by 2-bounded staleness consistency, as it
satisfies \texttt{SingleOrder} and \texttt{RealTime} with the sequential history
in Figure~\ref{fig:invalid_eg}, which is valid as per
\texttt{2-BoundedStalenessRVal}.

Similarly, sequential consistency relaxes \texttt{RealTime} guarantee of
linearizability to a \texttt{PRAM} guarantee which itself is composed of three
primitive session guarantees: \texttt{MonotonicReads}, \texttt{MonotonicWrites},
and \texttt{ReadYourWrite}~\cite{vukolic}. Again, the non-linearizable
history in Figure~\ref{fig:nonlin_eg} is allowed by sequential consistency, as
it satisfies \texttt{SingleOrder} and \texttt{PRAM} with the sequential history
in Figure~\ref{fig:seq_eg}, which is valid as per \texttt{RVal}.


Since it is often not obvious how to modify the WG algorithm to check other
consistency guarantees, most existing checkers have {\em limited versatility} as they
just check linearizability.  As discussed in
Section~\ref{sec:intro}, a Jepsen analysis of Zookeeper was not considered
meaningful because Knossos only checks for linearizability, and Zookeeper does
not guarantee linearizability. Zookeeper guarantees ordered sequential
consistency which is weaker than linearizability, but stronger than sequential
consistency~\cite{zk_osc}.

\subsection{Scalability}
The problem of determining whether a concurrent history is linearizable is known
to be NP-complete~\cite{npComp}. 
%
As discussed in Section~\ref{sec:intro}, Porcupine issue \#6 reports that
Porcupine hangs while checking a history with just 252
operations~\cite{porcupineIssue} since it has to consider many possible
permutations of writes. This problem was also acknowledged by the author of
Knossos linearizability checker in the same issue:

\begin{quote}
  We hit this in Knossos tests
  frequently, and while we've made as many efforts to optimize constant factors
  as possible, checking is still O(c!) with respect to concurrency--and
  concurrency almost always rises in real-world histories. You may be able to
  work around the issue via careful partitioning of your workload in both time
  and across distinct objects, reducing the state space, and performing frequent
  reads to prune the search space often.
\end{quote}

The suggested workarounds, such as performing frequent reads, are not always
feasible when checking real-world histories~\cite{existential_cons}.  For
instance, a shared log~\cite{lazylog} naturally have readers that lag
significantly after writers. Paritioning of the history may also not be 
possible, such as in a shared log with \code{append}/\code{get} interface.

Finally, reducing concurrency may not be desirable as bugs may appear only at
higher-levels of concurrency. For example, in our experiments with EPaxos,
out of the 3 bugs found, one appears in concurrency >10, rate >10 ops/s, and 
another appears in concurrency >10, rate >60 ops/s.

We discuss other consistency checkers like Facebook's~\cite{existential_cons} in
related work in Section~\ref{sec:related}.
\section{\name: A more versatile consistency checker}

\subsection{System-specific consistency checking}
The worst case runtime for strong consistency checkers is determined by the
total number of ways in which the operation history can be serialized. While
linearizability checkers already struggle to scale, as they must consider a large
number of possible operation permutations, checking
for weaker consistency guarantees will worsen scalability. This is because
weaker consistency guarantees have fewer ordering constraints on the operations.

To improve scalability, concurrently with this paper, the etcd authors suggest
in Porcupine Issue\#35~\cite{porcupineIssueEtcd} that the storage system can
provide {\em order hints} to reduce the number of possible operation
permutations, that the checker needs to check. In particular, they suggest that
etcd can respond to operations with a {\em revision number}, analogous to the
committed log index in Raft and the instance number in MultiPaxos. This revision
number can then be used to provide an additional ordering constraint (say
\code{etcd-Ordering}): \op{1}.revision < \op{2}.revision $\Rightarrow$ \op{1}
$\rb{S}$ \op{2}, \ie the operation with a lower revision number must get
serialized first.

We call this approach as {\em checking for system-specific
consistency guarantee}. Instead of checking for the basic consistency guarantee
(say linearizability), we are checking for ``etcd-linearizability'' which strengthens
linearizability to \code{SingleOrder} $\land$ \code{RVal} $\land$ \code{RealTime} 
$\land$ \code{etcd-Ordering}. The etcd authors observed significant speedups
when checking for etcd-linearizability, over checking for linearizability,
\ie without the \code{etcd-Ordering} constraint.

In addition to being faster and more scalable, checking for system-specific
consistency guarantees is clearly stronger than checking for the basic consistency
guarantee. For example, the operation history in Figure~\ref{fig:etcd_lin} is
linearizable, but not etcd-linearizable. This is because the equivalent
sequential history in Figure~\ref{fig:seq_eg} that helps satisfy
linearizability, does not preserve the \code{etcd-Ordering} constraint: \op{2}
has a {\em lower} revision number but is serialized {\em after} \op{1}. 

\tikzset{
    default tikz/.style={>=stealth, baseline=(current bounding box.center), scale=0.75},
}

\newcommand{\modInterval}[8][]{
    
    \draw[gray] (#2,#3) -- (#2+#4,#3);
    \draw[gray] (#2,#3-0.1) -- (#2,#3+0.1);
    \draw[gray] (#2+#4,#3-0.1) -- (#2+#4,#3+0.1);
    
    \node[above] at (#2+#4/2,#3) {#5};
    
    \if B#7
        \node[below] at (#2+#4*0.17647,#3-0.1) {#6};
    \else
        \if L#7
            \node[left] at (#2,#3) {#6};
        \else
            \node[right] at (#2+#4,#3) {#6};
        \fi
    \fi
    
    \if\relax\detokenize{#1}\relax
    \else
        \draw[gray] (#2+#4*#1,#3) node {$\bullet$};
    \fi

	\if\relax\detokenize{#8}\relax
    \else
        \node[right] at (#2+#4*1,#3+0.1) {#8};
    \fi
}

\newcommand{\intervalx}[8][]{%
    \interval[#1]{#2}{#3}{#4}{#5}{#6}{#7}%
    \if\relax\detokenize{#8}\relax
    \else
        \node[right] at (#2+#4*1,#3+0.1) {#8};%
    \fi
}

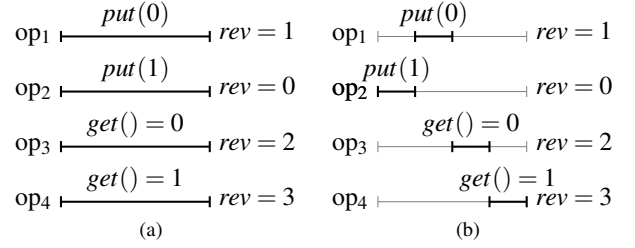
\begin{figure}[t]
  	\centering
	
	\resizebox{\columnwidth}{!}{%
	\def\k{2.7}
	\subfigure[\label{fig:etcd_lin}]{
		\begin{tikzpicture}[default tikz]
			\intervalx{0}{3}{\k}{$put(0)$}{\op{1}}{L}{$rev=1$}		
			\intervalx{0}{2}{\k}{$put(1)$}{\op{2}}{L}{$rev=0$}		
			\intervalx{0}{1}{\k}{$get() = 0$}{\op{3}}{L}{$rev=2$}	
			\intervalx{0}{0}{\k}{$get() = 1$}{\op{4}}{L}{$rev=3$}	
		\end{tikzpicture}
	}
	\subfigure[\label{fig:etcd_adjusted}]{
		\begin{tikzpicture}[default tikz]
			\def\x{\k/4}
			\modInterval{0}{3}{\k}{}{\op{1}}{L}{$rev=1$}	
			\modInterval{0}{2}{\k}{}{\op{2}}{L}{$rev=0$}	
			\modInterval{0}{1}{\k}{}{\op{3}}{L}{$rev=2$}	
			\modInterval{0}{0}{\k}{}{\op{4}}{L}{$rev=3$}	

			\interval{\x}{3}{\x}{$put(0)$}{}{L}			
			\interval{0}{2}{\x}{$put(1)$}{\op{2}}{L}	
			\interval{2*\x}{1}{\x}{$get() = 0$}{}{L}	
			\interval{3*\x}{0}{\x}{$get() = 1$}{}{L}	
		\end{tikzpicture}
	}
	}
	\label{fig:etcd_his}
    \vspace{-10pt}
	\caption{
	(a) A linearizable but not etcd-linearizable history.
	(b) Adjusted input history.}
    \vspace{-15pt}
\end{figure}

In our experiments in Section~\ref{sec:eval}, we indeed found that
implementation bugs can lead to histories that satisfy basic consistency
guarantees, but not system-specific consistency guarantees. We had to change the
experiment setup to make the same bug exhibit operation histories that also
violate the basic consistency guarantees. This indicates that checking for the
stronger system-specific consistency guarantees is helpful in reducing false
negatives.

The etcd authors realize etcd-linearizability checker via Porcupine by
modifying the input history. For each operation in the original history, either the
invocation time is delayed and/or the response time is advanced to generate the 
modified input history. For example, the operation history in
Figure~\ref{fig:etcd_lin} is modified using each operation's order hint (revision number) to the
one in Figure~\ref{fig:etcd_adjusted}\footnote{Because we are only delaying
invocation times and advancing response times, it can be seen that
linearizable points in the modified history will also be valid in the original
history, but not vice-versa. In other words, if the modified history is
linearizable, then the input history must be linearizable. It can also be seen 
that the modified history can only {\em reduce} the number of operation
permutations that may need to be tried for checking linearizability, leading to
faster and more scalable checking.}.

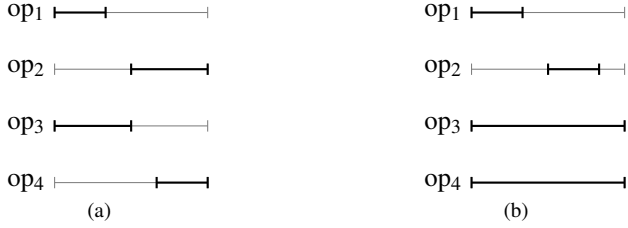
\begin{figure}[t]
  	\centering
	
	\def\k{2.7}
	\subfigure[\label{fig:mod_lost_order}]{
		\begin{tikzpicture}[default tikz]
			\def\x{\k/6}
			\modInterval{0}{3}{\k}{}{\op{1}}{L}{}	
			\modInterval{0}{2}{\k}{}{\op{2}}{L}{}	
			\modInterval{0}{1}{\k}{}{\op{3}}{L}{}	
			\modInterval{0}{0}{\k}{}{\op{4}}{L}{}	

			\interval{0}{3}{2*\x}{}{}{L}			
			\interval{3*\x}{2}{3*\x}{}{}{L}	
			\interval{0}{1}{3*\x}{}{}{L}	
			\interval{4*\x}{0}{2*\x}{}{}{L}	
		\end{tikzpicture}
	}\hfill
    \subfigure[\label{fig:mod_added_order}]{
		\begin{tikzpicture}[default tikz]
			\def\x{\k/6}
			\modInterval{0}{3}{\k}{}{\op{1}}{L}{}	
			\modInterval{0}{2}{\k}{}{\op{2}}{L}{}	
			\modInterval{0}{1}{\k}{}{\op{3}}{L}{}	
			\modInterval{0}{0}{\k}{}{\op{4}}{L}{}	

			\interval{0}{3}{2*\x}{}{}{L}	
			\interval{3*\x}{2}{2*\x}{}{}{L}	
			\interval{0}{1}{6*\x}{}{}{L}	
			\interval{0}{0}{6*\x}{}{}{L}	
		\end{tikzpicture}
	}
	
    \vspace{-10pt}
	\caption{Failed attempts to modify the history in Figure~\ref{fig:etcd_lin}
    with two partial ordering constraints: \op{1} before \op{2}, and \op{3} before \op{4}.
	(a) A modified history that adds a superfluous return-before order: \op{1} before \op{4}.
	(b) A modified history that loses \op{3} before \op{4} order, from the given partial 
    ordering constraints.}
	\label{fig:mod_his}
    \vspace{-15pt}
\end{figure}

This approach works well for etcd because the ordering hints, based on revision
numbers, form a {\em total order}.  However, several systems, like EPaxos,
can only provide {\em partial order
constraints} on their operations. A 2+2 partial order  {\em cannot} be converted
into a modified input history without losing some ordering constraints, as the
return-before ordering in operation histories can only express {\em interval
orders}. For example, it is not possible to adjust the history in
Figure~\ref{fig:etcd_lin}, with \op{1} before \op{2} and \op{3} before \op{4}
ordering constraints, without losing an ordering constraint and without 
introducing unintended orderings, as illustrated in Figure~\ref{fig:mod_his}.

For some systems, such as LazyLog, we will only be able to provide {\em soft
ordering constraints}, \ie in most cases, the operations will complete in the
given order but in rare situations, like failovers or errors, they may complete
in another order. 

We now describe a more versatile algorithm that can check different consistency
guarantees, such as sequential consistency, K-bounded staleness
consistency~\cite{cosmosdb}, ordered sequential consistency~\cite{zk_osc}, and
regular sequential consistency~\cite{sosp21_rsc}.  The algorithm can check for
system-specific strong consistency where system provides additional ordering
hints. The algorithm allows both partial and soft ordering constraints.


\subsection{A more general algorithm for checking consistency}
\begin{lstlisting}[language=Go, float=h, belowcaptionskip=-20pt,
	caption={A consistency model that allows K-bounded stale reads.},
	label=lst:valid]
kBoundedReadsConsistency := pine.Consistency{
  Oracles: {RealTime}
  Valid: {
    Init: func(model Model) interface() {
      return []int{model.Init()}
    }
    Step: func(states, input, output interface{},
        model Model) (bool, interface{}) {
      regInput := input.(registerInput)
      pastStates := states.([]int)
      if regInput.op { // get
        // Check if read is valid against 
        // any of last K states
        for _, s := range pastStates {
          ok, _ := model.Step(s, input, output)
          if ok {
            return ok, pastStates
          }
        }
        return false, pastStates
      } else { // put
        l := len(pastStates)
        ok, newState := model.Step(pastStates[l-1], input,
            output)
        // Truncate to K-1 states
        if l >= K {
          pastStates = pastStates[l-K+1:]
        }
        states = append(pastStates, newState.(int))
        return ok, states
      }
    },
  }
}
\end{lstlisting}

To develop a general algorithm, we observe that the WG algorithm checks all the
three primitive guarantees, required by linearizability as follows:

\begin{enum}
  \item \texttt{SingleOrder}: The final stack order determines the total order
  of the operations.
  \item \texttt{RVal}: While pushing every operation in the stack, the algorithm
  checks whether the sequential specification (Listing~\ref{lst:porcupine_reg})
  allows it on the current state \eg did this read operation return the latest
  write in the current state, obtained from applying operations in the stack. If
  the sequential specification does not allow it, the algorithm backtracks by
  popping operation(s) from the stack.
  \item \texttt{RealTime}: The algorithm does not move its \texttt{entry}
  pointer on the doubly linked list beyond the return of an operation, until the
  operation is linearized on the stack.
\end{enum}

\tikzset{
    default tikz/.style={node distance=0cm, scale=0.65},
    bx/.style={thick, draw, rectangle, minimum width=0.5cm, minimum height=0.5cm},
    op/.style={minimum size=0.6cm, inner sep=0pt, font=\small},
    stackbox/.style={bx, font=\small, minimum height=0.3cm, minimum width=1cm,fill=cyan!8},
    frontier/.style={draw,rectangle, rounded corners=8pt,inner sep=-1pt, red},
    node/.style={draw,circle,fill=white,op,minimum size=0.4cm}
}

\newcommand{\opc}[3]{
    \node[draw, thick,circle, minimum size=0.7cm, inner sep=0pt] (#1) at (#2,#3) {#1};
}
\newcommand{\rt}[2]{
    \draw[-stealth] (#1) -- (#2);
}
\newcommand{\tp}[2]{
    \draw[-stealth, thick, blue] ([xshift=-2cm,yshift=0.4cm]#1) -- ([xshift=2cm,yshift=0.5cm]#2);
}

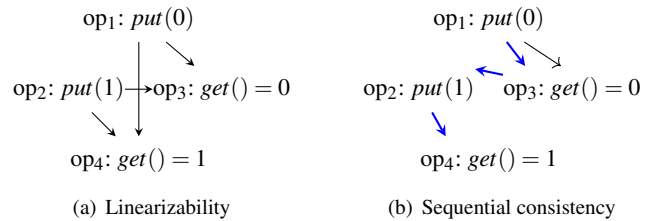
\begin{figure}[b]
    \centering
    \vspace{-13pt}
    \subfigure[\label{fig:lin_dag} Linearizability]{
        \begin{tikzpicture}[default tikz]
            \def\x{0.06cm}
            \def\y{0.5cm}
            \node (center) at (-1.5,1.8) {};
            \node (op1) [op,above=\y of center] {\op{1}: $put(0)$};
            \node (op2) [op,left=\x of center]  {\op{2}: $put(1)$};
            \node (op4) [op,below=\y of center] {\op{4}: $get()=1$};
            \node (op3) [op,right=\x of center] {\op{3}: $get()=0$};
            \rt{op1}{op4}
            \rt{op1}{op3}
            \rt{op2}{op3}
            \rt{op2}{op4}
        \end{tikzpicture}
    } \hfill
    \subfigure[\label{fig:seq_dag} Sequential consistency]{
        \begin{tikzpicture}[default tikz]
            \def\x{0.06cm}
            \def\y{0.5cm}
            \node (center) at (-1.5,1.8) {};
            \node (op1) [op,above=\y of center] {\op{1}: $put(0)$};
            \node (op2) [op,left=\x of center]  {\op{2}: $put(1)$};
            \node (op4) [op,below=\y of center] {\op{4}: $get()=1$};
            \node (op3) [op,right=\x of center] {\op{3}: $get()=0$};
            \draw[->] ([xshift=1cm]op1) -- ([xshift=2cm]op3);
            \tp{op1}{op3}
            \tp{op3}{op2}
            \tp{op2}{op4}
        \end{tikzpicture}
    }


            
    
    \vspace{-10pt}
    \caption{Operation DAGs for checking if the operation history 
    in Figure~\ref{fig:nonlin_eg} is allowed by linearizability or sequentially
    consistency respectively. In (b), we can find a topological order
    (highlighted by blue arrows), indicating that the history is allowed by
    sequential consistency.}
    \label{fig:dag}
\end{figure}
Our proposed algorithm continues to require the \texttt{SingleOrder} guarantee,
but it generalizes \texttt{RVal} to a \texttt{Validity} guarantee (\eg
\texttt{K-BoundedStalenessRVal}) and \texttt{RealTime} to an \texttt{Ordering}
guarantee that must be stronger than \texttt{PRAM} (\eg \texttt{RealTime}).  The
basic structure of our proposed algorithm is similar to the WG algorithm. We
maintain a stack of currently serialized operations, and backtrack when no
further operations can be pushed on the stack.

For a more general \texttt{Validity} checking, instead of passing just the
current state to the sequential specification (\eg to the \code{Step} function
in Listing~\ref{lst:porcupine_reg}), we pass a list of all the states from the 
current serialization. For example, Listing~\ref{lst:valid} defines a
\texttt{kBoundedReadsConsistency} that takes in a model such as
\code{registerModel} from the sequential specification of the register in
Listing~\ref{lst:porcupine_reg} to check for K-bounded stale reads in its
\code{Valid} method.

\begin{figure*}[t]
\centering
\begin{tikzpicture}[node distance=0.5cm and 0.5cm]
	\node[system] (c1) {};
	\node[system, anchor=south, xshift=1mm, yshift=-1mm](c2) at (c1.south) {};
	\node[system, anchor=south, xshift=1mm, yshift=-1mm](c3) at (c2.south) {\textbf{Clients}};

\node[system, align=left, right=1cm of c3] (storage) {
	\textbf{System implementation}\\
	\code{def handle(op: Op):}\\
	\code{\;\;  idx+=1}\\
	\code{\;\;  ...}\\
	\code{\;\;  return \{}\\
	\code{\;\;\;\;  val: op.ret,}\\
	\code{\;\;\;\;  \hint{\circled{a}~hint: idx}}\\
	\code{\;\;  \}}
};

	\node[trace, right=of storage, yshift=0.4cm] (t1) {};
	\node[trace, anchor=south, xshift=1mm, yshift=-1mm](t2) at (t1.south) {};
	\node[trace, align=left, anchor=south, xshift=1mm, yshift=-1mm](history) at (t2.south) {\circled{2}~\textbf{Client Traces}\\
	call(op1): \code{put(1)}, ts: 29\\
	ret(op1): \code{ok}, ts: 30, \hint{hint=17}\\
	call(op4): \code{get()}, ts: 33\\
	ret(op4): \code{0}, ts: 36, \hint{hint=20}
};
\node[system, right=of history, yshift=20pt] (parser) {
	\hint{Modified}\\\textbf{Trace parser}
};
\node[box, right=1.6cm of parser] (checker) {\textbf{Consistency}\\\textbf{Checker}};

\node[system, below=of parser] (model) {\textbf{Sequential}\\\textbf{Specification}};
\node[box, right=of model] (valid) {\textbf{Validity}\\\textbf{Checker}};
\node[system, right=of valid] (oracle) {
	\textbf{Ordering oracles}\\
	\hint{\circled{b}~$\prec(hint1, hint2)$}
};
\begin{scope}[on background layer]
\node[system, fit=(oracle) (valid) (model), inner sep=4pt, label={[txt, yshift=1pt]south:{\textbf{\hint{\circled{c}} System-specific consistency}}}] {};
\end{scope}

\node[txt, right=of checker] (result) {\circled{4}~\textbf{Fail/}\\\textbf{Success}};

\draw[arrow] (c3) -- node[midway, above, txt] {\circled{1}~Ops}  (storage);
\draw[thick] (storage) -- (history);
\draw[arrow] (history) -- (parser);
\draw[arrow] (parser) -- (checker);
\draw[arrow] (oracle) -- node[midway, right, txt] {\circled{3}~Build DAG}(checker);
\draw[arrow] (checker) -- (oracle);
\draw[arrow] (model) -- (valid);
\draw[arrow] (valid) -- (checker);
\draw[arrow] (checker) -- (result);
\end{tikzpicture}
\vspace{-10pt}
\caption{System-specific consistency checking with ordering hints.}
\label{fig:lin-arch}
\vspace{-10pt}
\end{figure*}
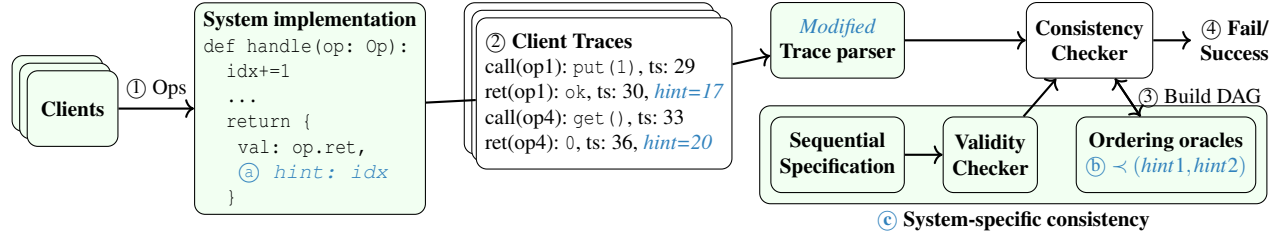

\texttt{Consistency} guarantees can declare \texttt{Oracle}s to express more
general \texttt{Ordering} guarantees. Each \texttt{Oracle} can take two
operations \op{1} and \op{2} and return one of 5 ordering outcomes: (1) there is
no ordering between \op{1} and \op{2}, (2) \op{1} must precede \op{2}, which we
call a hard constraint: \op{1}$\hard$\op{2}, (3) \op{2}$\hard$\op{1}, (4) \op{1}
should usually precede \op{2} in normal circumstances, but in case of fail overs
\etc it may not, which we call a soft constraint: \op{1}$\soft$\op{2}, and (5)
\op{2}$\soft$\op{1}. For example, Listing~\ref{lst:valid} declares
\code{RealTime} ordering \code{Oracle} which returns \op{1} $\hard$ \op{2} iff \op{1}
$\rb{H}$ \op{2}.

To check consistency, we maintain the operation history as a DAG instead of as a
doubly linked list in the WG algorithm. Each operation is a node in the DAG
with hard edges corresponding to $\hard$ and soft edges corresponding to
$\soft$ orders, as returned by the \code{Oracle}.  With these changes, the
consistency checking boils down to finding a topological order of the input DAG
over its hard edges such that the order is valid, as per the \texttt{Validity}
guarantee. We modify the standard topological order finding
algorithm to prioritize picking operations within
the current frontier using the soft edges.

Figure~\ref{fig:lin_dag} shows the input DAG for the history in
Figure~\ref{fig:nonlin_eg} when we check it for linearizability, and
Figure~\ref{fig:seq_dag} shows the input DAG for the same history when we check
it for sequential consistency.  When using the \texttt{RVal} validity criteria,
we can find a valid topological order for the DAG in Figure~\ref{fig:seq_dag}
(highlighted by blue arrows), but not for the DAG in Figure~\ref{fig:lin_dag},
indicating that the history is allowed by sequential consistency, but not by
linearizability.

A naive implementation materializes the full operation DAG, which requires
checking ordering hints between every pair of operations. We discuss our
optimized algorithm in Section~\ref{sec:hintalgo}.

\subsection{System Design}
Figure~\ref{fig:lin-arch} shows how \name modifies the consistency checking
workflow of Porcupine and Knossos. All the system-specific components are
highlighted in green (ignore \hint{blue italics text} for now).  A testing
framework like Jepsen~\cite{jepsen} spawns storage system nodes and lets clients
\circled{1}~send operations to the system.  The clients \circled{2}~record a
trace of all the operations. This trace captures the timestamps of calls and
returns for every operation. 
The trace is sent to the system-specific trace parser which \circled{3}~helps
build the operation history and sends it to the consistency checker. The checker
\circled{4}~checks the operation history to output if the input history is
allowed by the consistency guarantee.

To check system-specific consistency guarantee, we modify this workflow as
follows:

\hint{\circled{a}}~The system implementation returns an additional ordering
\hint{hint} when returning each operation (\eg revision number for etcd, zxid
for ZooKeeper). 

\hint{\circled{b}} The system-specific trace parser is modified to process 
traces augmented with ordering hints. It then uses the operation hints to expose
additional ordering constraints between operations through a custom
\hint{oracle}.  As discussed already, for any two operations \op{1} and \op{2},
the oracle can return 5 types of orders: no ordering, \op{1}$\hard$\op{2},
\op{2}$\hard$\op{1}, \op{1}$\soft$\op{2}, and \op{2}$\soft$\op{1}. For example,
the Zookeeper~\cite{zookeeper} oracle returns \op{1} $\hard$ \op{2} 
when \op{1}$\!\!$'s zxid is lesser than that of \op{2}. When the two zxid are equal,
\op{1} $\hard$ \op{2} if \op{1} is a write and \op{2} is a read.

\hint{\circled{c}} The system specifies a \hint{system-specific consistency}
containing the \code{Validity} specification (\eg K-bounded staleness validity)
and a list of \code{Ordering} oracles. For example, Zookeeper-specific
consistency declares the ZooKeeper oracle described above, and a
\code{RealTimeWW} oracle for ordered sequential consistency which only applies
return-before real time orders on writes.

The consistency checker repeatedly invokes the oracles to build the
operation DAG (such as in Figure~\ref{fig:lin_dag}), and then finds a valid
topological order in this DAG using the validity specification.

\begin{table*}[t!]
\centering
{\scriptsize
\begin{tabular}{|c|c|c|c|l|}
\hline
\textbf{System} & \textbf{Impl.} & \textbf{\hint{\circled{a}} LoC} & \textbf{\hint{\circled{b}} LoC} & \textbf{System-specific ordering oracle} \\
\hline

\multicolumn{5}{c}{Eager total ordering linearizable systems} \\\hline


HoliPaxos~\cite{holipaxos} & Go & 3 of 3.1K & 28 & 
	\op{1} $\hard$ \op{2}: \op{1}.instance < \op{2}.instance 
	\\ \hline

etcd~\cite{etcd} & Go & 0 of 151K & 24 & 
	\op{1} $\hard$ \op{2}: \op{1}.revision < \op{2}.revision 
	\\ \hline

Redis Raft~\cite{redisraft} & C & 30 of 28K & 30 & 
	\makecell[l]{
        \wop{1} $\hard$ \wop{2}: \wop{1}.logidx < \wop{2}.logidx \\ 
        \rop{1} $\hard$ \rop{2}: \rop{1}.last-applied-instance < \rop{2}.last-applied-instance \\ 
		r $\hard$ w: r.last-applied-instance < w.logidx \\
		w $\hard$ r: w.logidx $\leq$ r.last-applied-instance
    } \\ \hline

\multicolumn{5}{c}{Dependency tracking linearizable systems} \\\hline

EPaxos~\cite{epaxos} & Go & 42 of 7.3K & 55 & 
	\makecell[l]{
	\op{1} $\hard$ \op{2}:  \op{1}.inst $\le$ \op{2}.deps[\op{1}.replica] $\land$ \op{2}.inst $\not\le$ \op{1}.deps[\op{2}.replica] \\
	\op{1} $\hard$ \op{2}:  \op{1}.inst $\le$ \op{2}.deps[\op{1}.replica] $\land$ \op{2}.inst $\le$ \op{1}.deps[\op{2}.replica] \\
	\>\>\>\> $\land$ \op{1}.seq < \op{2}.seq \\
	}
	\\ \hline

EPaxos revisited~\cite{epaxos_revisited} & Go & 67 of 5.5K & 60 & 
	\makecell[l]{
	In addition to above, \\
	\op{1} $\soft$ \op{2}: \op{1}.timestamp < \op{2}.timestamp \\
	}
	\\ \hline

\multicolumn{5}{c}{Lazy ordering linearizable systems} \\\hline

Erwin~\cite{lazylog} & C++ & 20 of 4.8K & 80 & 
	\makecell[l]{
		\wop{1} $\hard$ \wop{2}: \wop{1}.view $<$ \wop{2}.view \\
		\wop{1} $\hard$ \wop{2}: $\forall \text{s} \in$ Sequencers: \wop{1}.seqIdx[s] $<$ \wop{2}.seqIdx[s]\\
		w $\hard$ r: $\forall \text{s} \in$ Sequencers: w.seqIdx[s] $\le$ r.idx \\
		\wop{1} $\soft$ \wop{2}: \wop{1}.seqIdx[leader] < \wop{2}.seqIdx[leader] \\
		w $\soft$ r: w.seqIdx[leader] $\le$ r.idx \\
	}
	\\ \hline

Skyros~\cite{nilext} & C++ & 42 of 5.6K & 60 & 
	\makecell[l]{
		\op{1} $\hard$ \op{2}: \op{1}.view < \op{2}.view \\
		\wop{1} $\hard$ \wop{2}: \wop{1}.DIdx[leader] < \wop{2}.DIdx[leader] $\land$ \\
		\>\>$\exists\; \text{Supermajority} \subseteq \text{Sequencers}:$
		\>\>\>\>$\forall \text{s} \in$ Supermajority: \wop{1}.DIdx[s] $<$ \wop{2}.DIdx[s]\\
		w $\hard$ r: w.DIdx[leader] $\le$ r.CIdx[leader] \\
		\rop{1} $\hard$ \rop{2}: \rop{1}.CIdx[leader] $\le$ \rop{2}.CIdx[leader] \\
		\wop{1} $\soft$ \wop{2}: \wop{1}.DIdx[leader] < \wop{2}.DIdx[leader] \\
	}
	\\ \hline

\multicolumn{5}{c}{Non-linearizable systems} \\\hline

Zookeeper~\cite{zookeeper} & Java & 0 of 127K & 30 & 
	\makecell[l]{
		\op{1} $\hard$ \op{2}: \op{1}.zxid < \op{2}.zxid  \\
		w $\hard$ r: w.zxid = r.zxid 
	}
	\\ \hline

Gryff-RSC~\cite{gryff} & Go & 30 of 18K & 18 & 
	\op{1} $\hard$ \op{2}: \op{1}.carstamp <  \op{2}.carstamp
	\\ \hline


\end{tabular}
}
\caption{System implementations and lines of changes to \hint{\circled{a}}~add
ordering hints to the implementation, and to \hint{\circled{b}}~incorporate
ordering hints in the system's trace parser and ordering oracle. Ordering
constraints are provided for single key operations. $w$ represents a write
operation, $r$ represents a read operation, and $op$ represents any of the two.
\\For HoliPaxos, etcd, Redis Raft, and ZooKeeper the ordering constraints order
operations by their position in the consensus log. Redis Raft and Zookeeper do 
not put reads in the consensus log. So, we order reads after writes when they
return with the same last-applied-instance and zxid respectively. For Gryff-RSC,
we order operations by their carstamps.
\\For EPaxos, we use the
ordering constraint proved by Theorem 5 from the EPaxos proof of
correctness~\cite{epaxos}, which says that if there is a cycle of dependencies,
then the operation with the lower sequence number executes first. If there are
no cycles, then the dependent operation executes next. Note that all
operations conflict for operations to single key. Also note that since sequence
numbers can be equal, the order constraint only defines a partial order.
EPaxos revisited adds timestamp-ordered queueing. We return this timestamp with
each operation and use it for soft order constraints.
\\For Erwin, we use the following hard order constraints: writes that completed
in an earlier view happens first, and writes that had lower sequence number from
all the sequencers happen first. If a leader does not fail, its sequence number 
determines the order in which writes are applied. After a failover, writes might
get applied in a different order. We therefore use leader's sequence number only
in soft order constraints.
\\For Skyros, we use the following hard-order constraints: 
operations that complete in an earlier view must precede those that 
complete in a later view. The index in the \textit{durability log} orders writes, 
while the index in the \textit{consensus log} orders reads. 
If supermajority of sequencers doesn't agree on the order of writes from the
leader's durability log, then the writes might be applied in a different 
order during failover. We therefore relax this supermajority condition 
and use the resulting ordering as a soft-order constraint.
}
\label{tab:protocol_impl}
\end{table*}

Clients and sequential specification remain unmodified. Our proposed
modifications are lightweight. Table~\ref{tab:protocol_impl} shows that we can
check the system-specific ordering oracles that we add for \numsut systems.  We
could add the order hints in less than \hintloc lines of code for all the
systems.
\subsection{Incorporating ordering hints to the operation DAG}
\label{sec:hintalgo}

The consistency model enforces ordering constraints on pair of operations, which
can be represented as a DAG with operations as nodes and constraints as edges. A
brute-force construction requires $O(n^2)$ pairwise comparisons, which might be
infeasable for large histories. So, we simplify the DAG construction by
enforcing that the \texttt{Ordering} guarantee should be stronger than the
\texttt{SessionOrder} guarantee: operations from a client must be ordered in the
same order as they were issued. So, the DAG is simplified to $c$ chains, one per
client. There will be a few cross-client dependencies. We will see that we can
represent the cross-client dependencies for each operation with only two edges
(one incoming and one outgoing) to each of the other clients.

Consider any operation $\text{op}_n^i$, the $n^{th}$ operation from client $i$.
For every other clients $j$, there exists a maximal range $[u, v]$ of operations
from client j are concurrent with $\text{op}_n^i$. Operations before this range
have outgoing dependencies to $\text{op}_n^i$ and operations after this range
have incoming dependencies from $\text{op}_n^i$. We record the start of this
range as $\text{op}_n^i.\text{start}[j] = u$. We will see why this is helpful
but first, we show how can we compute this faster by making use of total order
and transitivity.

Suppose $\text{op}_n^i.start[j] = u$, so for all $m < u$, the dependency
$\text{op}_m^j \rightarrow \text{op}_n^i$ holds. By \texttt{SessionOrder}, we
get the dependency $\text{op}_n^i \rightarrow \text{op}_{n+1}^i$. And by
transitivity we get $\text{op}_{m}^j \rightarrow \text{op}_{n+1}^i$. Therefore
$\text{op}_{n+1}^i.start[j] \ge u$. So, we know a lower bound for
$\text{op}_{n+1}^i.start[j]$. We need to search only among the operations in
client $j$ in the range $[u, \text{\#Ops\;from\;client\;j}]$. Since this is a
totally ordered set, we can perform binary search on this range. Instead of the
standard binary search, we perform exponential search starting from $u$. This
performs better when the range is small.

To check any arbitrary consistency model, we first construct the chains and then
construct the DAG. Each operation $\text{op}$ contains a boolean attribute
$\text{op}.visited$ which denotes whether $\text{op}.start$ has been computed.
We initialize $op_0^i.start$ for each client $i$.

To find the single order, the checker explores all topological orderings of the
DAG. It maintains a frontier, the set of operations with no non-serialized
incoming dependencies. This can be easily computed using the \texttt{start}
attribute of an operation. At each step, it picks an operation from the frontier
and attempts to serialize it against the current model state. The current
serialization is stored in a stack which is used for backtracking. The
operations from the frontier are explored in the soft-hint order. Within the
frontier, the checker compares all pairs of operations to discover the soft
orders. This is possible because the size of frontier is at most $O(c)$. When no
operation from the frontier is serializable, the checker backtracks, popping an
operation from the stack.


%



\section{Evaluation}
\label{sec:eval}

We have added $\sim$900 lines of Go code to \texttt{Porcupine} to realize the ideas in
this paper. We call our implementation \name. In this section, we evaluate the
following research questions:

\begin{enumerate}
\item Is checking for system-specific consistency guarantees effective in
finding bugs which may not be caught otherwise? (Section~\ref{sec:eval-effective})
\item How do hard and soft order constraints influence the runtime and 
scalability of \name? (Section~\ref{sec:eval-ablation})
\item Does \name add memory and time overhead over the \texttt{Porcupine}
implementation? (Section~\ref{sec:overhead})
\end{enumerate}

%

\subsection{Evaluation Setup}
We follow a two-phase setup for our evaluations. The setup is summarized in
Table~\ref{tab:experiment-dimensions}.

In the first phase, we generate a comprehensive set of operation histories by
running a suite of experiments on various distributed storage systems. Each
storage system is modified to respond with operation hints, as was described in
Table~\ref{tab:protocol_impl}.  We deploy each system in Jepsen's~\cite{jepsen}
Docker cluster containing 5 nodes, isolated within a virtual Docker network, so
that Jepsen can inject faults while testing each system. We run each experiment for 2 
minutes, three times, where clients send read and write requests in a closed loop. 

In the second phase, we analyze these histories using \name, while varying
ordering oracles, as well as with \porcupine to detect consistency violations,
and to compare their runtime performance. Both implementations are written in Go.
All Experiments were performed on a system with an Intel Core i9-12900 CPU 
(24 logical threads), 32 GB RAM, and a 1TB SK hynix NVMe SSD, 
running Docker v28.1.1 with Intel VT-x enabled. 

For finding bugs, we run each storage system with different fault injection
scenarios using Jepsen. We use 5, 25, and 125 Jepsen concurrent clients with
varying read ratios to send requests. In our setup, the concurrent Jepsen
clients are realized by a single Jepsen control node which sends requests with
different client IDs. 

For measuring scalability of consistency checking, we do not use Jepsen clients 
as it gets bottlenecked by the single Jepsen control node, producing artificially
smaller operation histories. We instead use real client threads and run the 5
replicas of each system as 5 processes \ie not inside the Jepsen's docker
cluster, without any fault injection. We use 8, 12, 16, and 20 concurrent clients
with varying read ratios. For each experiment setting, we measure the end-to-end
time taken to output a consistency verdict and the peak memory consumption
during the checking process.  Across all graphs, a \textcolor{red}{x} mark
indicates that the checker timed out after 2 minutes and failed to produce a
verdict.  
We report averages across the three operation history traces for each setting.

For runtime performance of checkers, we only show results for operation
histories that were found to be consistent, as inconsistent operation histories
do not reflect the true runtime of the checkers. A checker can exit quickly on
an inconsistent history, if say it is unable to put just the first two
operations in a single order.  

\subsubsection{Setup limitations}
For Redis Raft and etcd, we could use their Jepsen clients and
server setups. For others, we faced several difficulties in running the systems
within Jepsen. The most notable one is Erwin. Instead of using hardware RDMA in 
their original implementation, we tried to use software RDMA (\texttt{rxe}).  In
this setup, Erwin
encountered RDMA read failures and
never progressed to stable end-to-end ordering and storage. In addition, we
observed an independent deadlock bug in the shard server's read path. 
Due to this, we are only able to report runtime performance of checkers with
Erwin where both the storage replicas and the clients are running outside
Jepsen's docker containers. Since Erwin realizes a totally ordered log with
append/get operations, it is much harder to linearize than the other systems
which we test with a shared register with put/get operations. Due to this, at
higher concurrency and lower read ratios, all checkers timeout with Erwin. We
therefore report results with 4, 6, 8, and 10 client threads only.

For Gryff-RSC, the client itself does additional bookkeeping of the last
observed value to send with the next operation. Instead of moving this client
into Jepsen, we send a read-write request from the Jepsen client to Gryff's
client which then forwards it to Gryff servers. 


We found consistency violation bugs in Gryff-RSC, EPaxos, EPaxos-revisited, and
HoliPaxos.  We were able to root cause and fix some consistency violations for
the last three systems, as we will describe in Section~\ref{sec:case_studies}.
EPaxos and EPaxos-revisited still had more consistency violations which we are
still investigating.


\begin{table}[t!]
	{\footnotesize
\centering
\begin{tabular}{|p{0.26\columnwidth}|p{0.60\columnwidth}|}
\hline
\textbf{Dimension} & \textbf{Values} \\
\hline
System &
HoliPaxos~\cite{holipaxos_imp}, RedisRaft~\cite{redisraft_imp}, 
etcd~\cite{etcd_imp}, EPaxos~\cite{epaxos_imp}, EPaxos-Revisited~\cite{epaxos_revisited_imp}, 
LazyLog~\cite{erwin_imp}, Zookeeper~\cite{zookeeper_imp}, 
Gryff-RSC~\cite{gryff_rsc_imp}, Skyros~\cite{nilext}
, YugabyteDB~\cite{yugabyte_imp}\\\hline

Read ratio & 0.01, 0.33, 0.66, 0.99 \\\hline

\multicolumn{2}{c}{Runtime performance testing} \\\hline
Client threads & 8, 12, 16, 20 \\\hline

\multicolumn{2}{c}{Failure-injection testing} \\\hline
Nemesis &
Noop, Clock scrambling, Partition (random half), Partition (split one), \\
& Hammer-time, Majority ring \\\hline

Client threads & 5, 25, 125 \\\hline

\end{tabular}
	}
\caption{Experiment setup configurations. Each experiment is repeated 3 times
and is run for 2 minutes using closed-loop clients. We deploy each storage
system with 5 replicas. For failure injection testing, the servers and clients
are deployed within Jepsen's docker cluster.}
\label{tab:experiment-dimensions}
    \vspace{-10pt}
\end{table}

\subsection{Effectiveness of system-specific consistency checking}
\label{sec:eval-effective}

System-specific consistency guarantees are {\em stronger} than basic consistency
guarantees. For example, ``etcd-linearizability" adds an {\em additional} ``etcd
ordering oracle" based on revision numbers. 

\paragraph{Evaluation highlights.} We found that checking for this stronger
guarantee can reduce false negatives and speed up consistency checking as there
are fewer possible topological orders to explore. We also found that the
operation hints can be a useful tool to debug consistency violations. 

We discuss the consistency violation bugs and their root causes in
Section~\ref{sec:case_studies} and compare checker performance in
Section~\ref{sec:perf}.

\subsubsection{Consistency violations}
\label{sec:case_studies}


\tikzset{
    default tikz/.style={>=stealth, baseline=(current bounding box.center), scale=0.75},
    op/.style={minimum size=0.7cm, inner sep=1pt, font=\small},
}

\newcommand{\intervalhint}[9][]{
    \interval[#1]{#2}{#3}{#4}{#5}{#6}{#7}
    \if B#9
        \node[below] at (#2+#4*0.5,#3-0.2) {#8};
    \else
        \node[right] at (#2+#4,#3) {#8};
    \fi
    \if\relax\detokenize{#8}\relax
    \else
    \fi
    \coordinate (#6-start) at (#2+0.2,#3);
    \coordinate (#6-end) at (#2+#4-0.2,#3);
    \coordinate (#6-mid1) at (#2+#4*0.5-0.2,#3);
    \coordinate (#6-mid2) at (#2+#4*0.5+0.2,#3);
}

\newcommand{\netpartition}[1]{
    \draw[dotted, thick] (#1,1.75) -- (#1,5);
    \node[right] at (#1,5) {\scriptsize network partition};
}

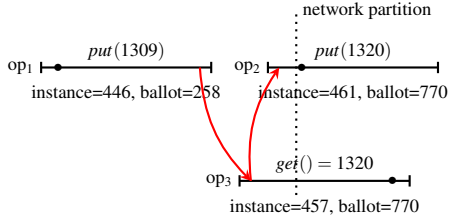
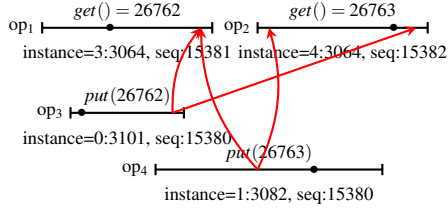
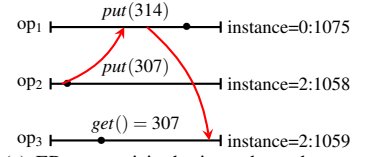
\begin{figure*}[t]
\centering
\subfigure[b][HoliPaxos does not handle operation retries in an idempotent manner,
after a new leader gets elected. For this history, \op{2} was executed three times,
at instance number 451, 456, and 461. \label{fig:holipaxos-bug}]{
    \centering
    \scriptsize
        \begin{tikzpicture}[default tikz]
            \intervalhint[0.1]{0}{4}{3}{$put(1309)$}{\op{1}}{L}{instance=446, ballot=258}{B}
            \intervalhint[0.2]{4}{4}{3}{$put(1320)$}{\op{2}}{L}{instance=461, ballot=770}{B}
            \intervalhint[0.9]{3.5}{2}{3}{$get() = 1320$}{\op{3}}{L}{instance=457, ballot=770}{B}
        \netpartition{4.5}
        \draw[->, thick, red] (\op{1}-end) to[bend right=20] (\op{3}-start);
        \draw[->, thick, red] (\op{3}-start) to[bend left=20] (\op{2}-start);
        \end{tikzpicture}
}\hfill
\subfigure[b][EPaxos replicas break dependency cycles in an inconsistent manner.
For this history, replicas 3 and 4 executed the write operations \op{3} and \op{4}
in different orders, leading to diverging reads.\label{fig:epaxos-bug1}]{
    \centering
    \scriptsize
        \begin{tikzpicture}[default tikz]
            \intervalhint[0.4]{0}{4}{3}{$get() = 26762$}{\op{1}}{L}{instance=3:3064, seq:15381}{B}
            \intervalhint[0.8]{3.8}{4}{3}{$get() = 26763$}{\op{2}}{L}{instance=4:3064, seq:15382}{B}
            \intervalhint[0.1]{0.5}{2.5}{2}{$put(26762)$}{\op{3}}{L}{instance=0:3101, seq:15380}{B}
            \intervalhint[0.7]{2}{1.5}{4}{$put(26763)$}{\op{4}}{L}{instance=1:3082, seq:15380}{B}
        \draw[->, thick, red] (\op{4}-mid1) to[bend left=20] (\op{1}-end);
        \draw[->, thick, red] (\op{4}-mid1) to[bend right=20] (\op{2}-start);
        \draw[->, thick, red] (\op{3}-end) to[bend left=20] (\op{1}-end);
        \draw[->, thick, red] (\op{3}-end) to[bend right=0] (\op{2}-end);
        \end{tikzpicture}
}\hfill
\subfigure[b][EPaxos-revisited misses dependent operations, when executing. In
this history, replica 2 returns a read operation \op{3}, before executing the
write operation \op{1}, on which \op{3} depended upon. \label{fig:epaxos-bug2}]{
    \centering
    \scriptsize
        \begin{tikzpicture}[default tikz]
            \intervalhint[0.8]{0}{4}{3}{$put(314)$}{\op{1}}{L}{instance=0:1075}{R}
            \intervalhint[0.1]{0}{3}{3}{$put(307)$}{\op{2}}{L}{instance=2:1058}{R}
            \intervalhint[0.3]{0}{2}{3}{$get() = 307$}{\op{3}}{L}{instance=2:1059}{R}
            \draw[->, thick, red] (\op{2}-start) to[bend right=20] (\op{1}-mid1);
            \draw[->, thick, red] (\op{1}-mid2) to[bend left=20] (\op{3}-end);
        \end{tikzpicture}
}
\vspace{-10pt}
\caption{Interesting operations of linearizable histories that violated
system-specific consistency guarantees. The linearization shown by the
linearization points is disallowed by the systems' ordering hints, shown by red
arrows.}
\end{figure*}
\paragraph{HoliPaxos.} When we tested HoliPaxos under network
partitions, we found histories that were linearizable but {\em not}
HoliPaxos-linearizable.  Figure~\ref{fig:holipaxos-bug} shows one such example.
The history can be linearized into the order \op{1}, then \op{2}, then \op{3}.
But this linearization is {\em not} allowed by HoliPaxos-linearizability,
because \op{2} has a higher instance number than \op{3} and therefore must
happen after \op{3}. In other words, \op{3} reads a value that is committed {\em
later} by the consensus protocol.

Upon investigating the replication log, we found that HoliPaxos committed the
operation \op{2} {\em thrice} at instance numbers 451, 456, and 461, with only
the last instance being returned to the client. In other words, HoliPaxos is not
idempotent, \ie it could apply the same operation multiple times, especially
after a new leader is elected (notice that \op{1}'s response contained
ballot=258, and \op{2}'s response contained ballot=770 indicating a
leader reelection after the partition).




To expose non-idempotent request handling, we changed \code{put} operations to
\code{increment} operations. 
With this modification, linearizability also failed confirming the bug.

We fix the idempotency issue by implementing the ideas from RIFL~\cite{RIFL} where 
clients generate a unique ID for each operation and the servers track and store
the result of each operation based on its ID. If the server receives an
operation with an ID that it has already executed, it simply returns the stored
result of the operation, without re-executing the operation. This fix eliminated
the consistency violations.

\paragraph{EPaxos-revisited.} We found three consistency violation bugs in
EPaxos-revisited. All the three bugs revealed themselves, even without any fault
injection. The first bug was also present in EPaxos. The last bug could also be
caught by linearizability checking, the other two only got caught when checking
for EPaxos-linearizability.

{\em 1. Inconsistent cycle breaking.} Figure~\ref{fig:epaxos-bug1} shows a
history that is linearizable, but {\em not} EPaxos-linearizable. The dependency
order hint returned for the operations show that both the reads \op{1} and
\op{2} must happen {\em after} the two writes \op{3} and \op{4} (indicated by
the red arrows). However, the operation \op{1} reads the value written by \op{3},
and the operation \op{2} reads the value written by \op{4}.

The order hints revealed the following important details which were useful in
root-causing this issue.  The two reads \op{1} and \op{2} have different command
leaders: replica 3 and replica 4 respectively. The two writes \op{3} and \op{4}
also had different command leaders, replicas 0 and 1 respectively.  The two
write operations formed a cyclic dependency, both having an equal sequence
number of 15380. Therefore, we suspected that the two replicas 3 and 4 which led
the two reads, did not execute the two writes in a consistent manner.

When EPaxos replicas track dependencies between operations, they can discover
that the operation dependencies form a cycle. Each replica independently breaks
the cycle by sorting the operations, and then executing them in this
sorted order. To realize replicated state machines, all replicas {\em must}
break dependency cycles in a consistent manner. However, this is where the
bug stems from. In the implementation, different replicas can sort the
operations with a dependency cycle in inconsistent orders.

The fix is shown in the below Listing, where the sorting comparator did
not handle the case where the sequence numbers of two operations can be
identical.

\begin{lstlisting}[language=go, escapechar=@, label=lst:cycle]
func (na nodeArray) Less(i, j int) bool {
@\textcolor{green!70!black}{+\ \ if na[i].Seq == na[j].Seq \{}@
@\textcolor{green!70!black}{+\ \ \ \ if na[i].ReplicaId == na[j].ReplicaId \{}@
@\textcolor{green!70!black}{+\ \ \ \ \ \ \ return na[i].InstanceNo < na[j].InstanceNo}@
@\textcolor{green!70!black}{+\ \ \ \ \}}@
@\textcolor{green!70!black}{+\ \ \ \ return na[i].ReplicaId < na[j].ReplicaId}@
@\textcolor{green!70!black}{+\ \ \}}@
   return na[i].Seq < na[j].Seq
}
\end{lstlisting}

{\em 2. Missed operations.} After fixing the previous bug, we again found
another history that is linearizable but {\em not} EPaxos-linearizable. The
interesting operations of the history are shown in Figure~\ref{fig:epaxos-bug2}.
The order hints reveal that the write operation \op{1} should happen before the
read operation \op{3}, but the read operation returned a stale write by \op{2}.

We could not reproduce this bug with EPaxos. Upon diffing the two source codes,
we found that when the execution thread is traversing the dependency graph of
operations, if the replica does not yet have the dependent operation or if it
has not yet committed the operation, EPaxos would sleep, waiting for the
operation to appear and for the commit to happen. EPaxos-revisited changed these
\code{sleep} to \code{return false} as shown in the below Listing.
However, when returning \code{false} from the \code{strongconnect} method, it is
mandatory to clear the visited \code{stack}, otherwise the operation gets
skipped. The fix was to clear the \code{stack}, whenever \code{strongconnect}
returns \code{false}.

\begin{lstlisting}[language=go, escapechar=@, label=lst:missed]
func (e *Exec) strongconnect(v *Instance, index *int) bool
{
  // ...
  for i := e.r.ExecedUpTo[q] + 1; i <= inst; i++ {
    for e.r.InstanceSpace[q][i] == nil || ... {
      // change from time.Sleep(1000 * 1000)
      return false
    }
    for e.r.InstanceSpace[q][i].Status != 
        epaxosproto.COMMITTED {
      // change from time.Sleep(1000 * 1000)
      return false
    }
    // ...
    if !e.strongconnect(w, index) {
      for j := l; j < len(stack); j++ {
        stack[j].Index = 0
      }
      stack = stack[0:l]
      return false
    }
  }
  // ...
}
\end{lstlisting}

{\em 3. Blind writes.} EPaxos-revisited adds \code{PUT\_BLIND} writes which
return after committing, but before being executed, to measure and contrast
commit latency with execution latency. However, the implementation missed
declaring that \code{PUT\_BLIND} conflicts with other writes to the same key.
The fix was to add this check as shown in the below Listing.

\begin{lstlisting}[language=go, escapechar=@, label=lst:putblind]
// src/state/state.go
func Conflict(gamma *Command, delta *Command) bool {
  if gamma.K == delta.K {
    if gamma.Op == PUT || delta.Op == PUT 
@\textcolor{green!70!black}{+\ \ \ \ \ \ \ || gamma.Op == PUT\_BLIND || delta.Op == PUT\_BLIND \{}@
      return true
    }
  }
}
\end{lstlisting}

{\em Takeaway: System-specific consistency checking reduces false-negatives, as
they are stronger guarantees than the basic consistency guarantees. The order
hints can assist the developers in root-causing the reason for the consistency
violation.}

\subsubsection{Checker performance}
\label{sec:perf}
\begin{figure*}[!ht]
\begin{minipage}[t]{0.725\textwidth}
    \centering
    \pgfmathsetmacro{\total}{0.98}
    \pgfmathsetmacro{\leftfrac}{0.525}
    \pgfmathsetmacro{\rightfrac}{\total-\leftfrac}

    \includegraphics[width=\leftfrac\linewidth]{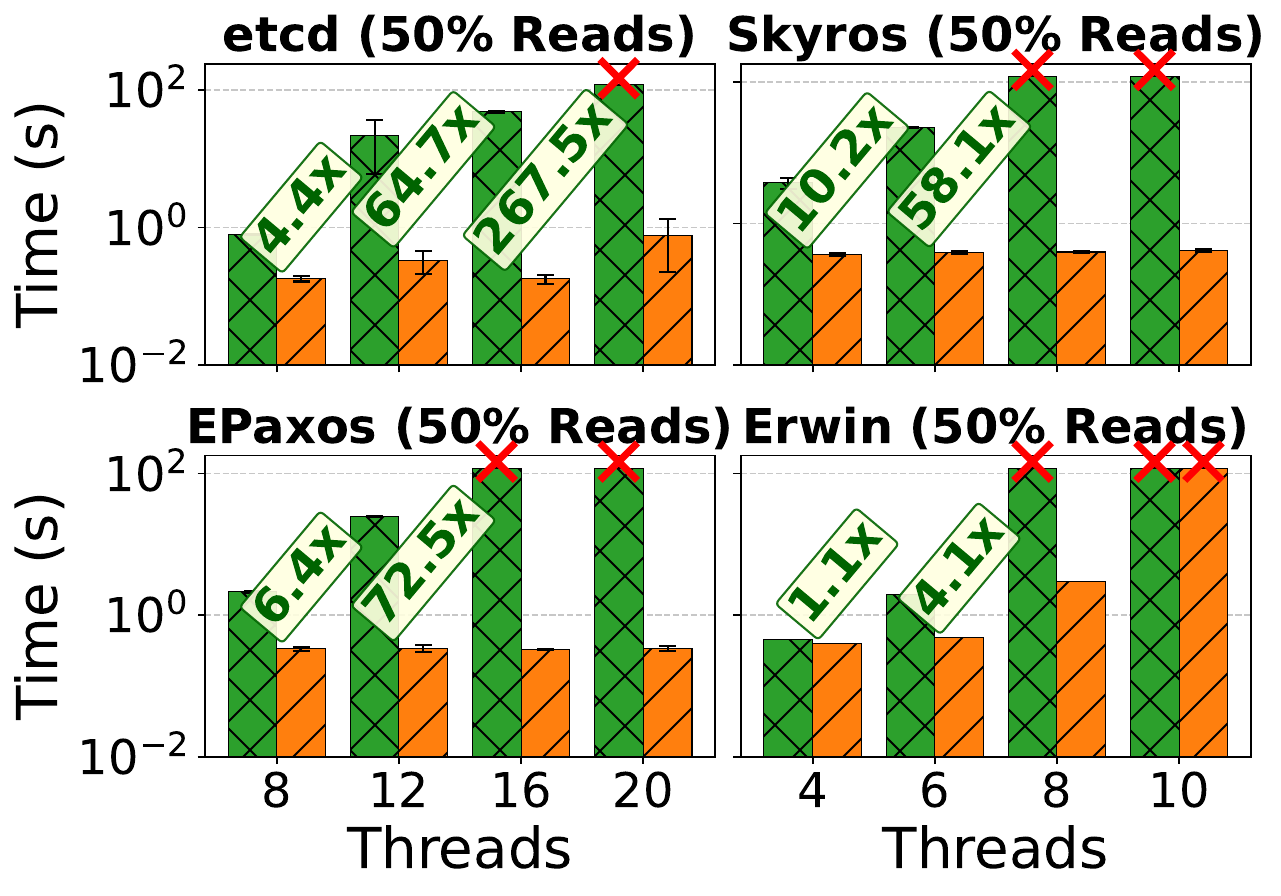}
    \hspace{-7pt}
    \includegraphics[width=\rightfrac\linewidth]{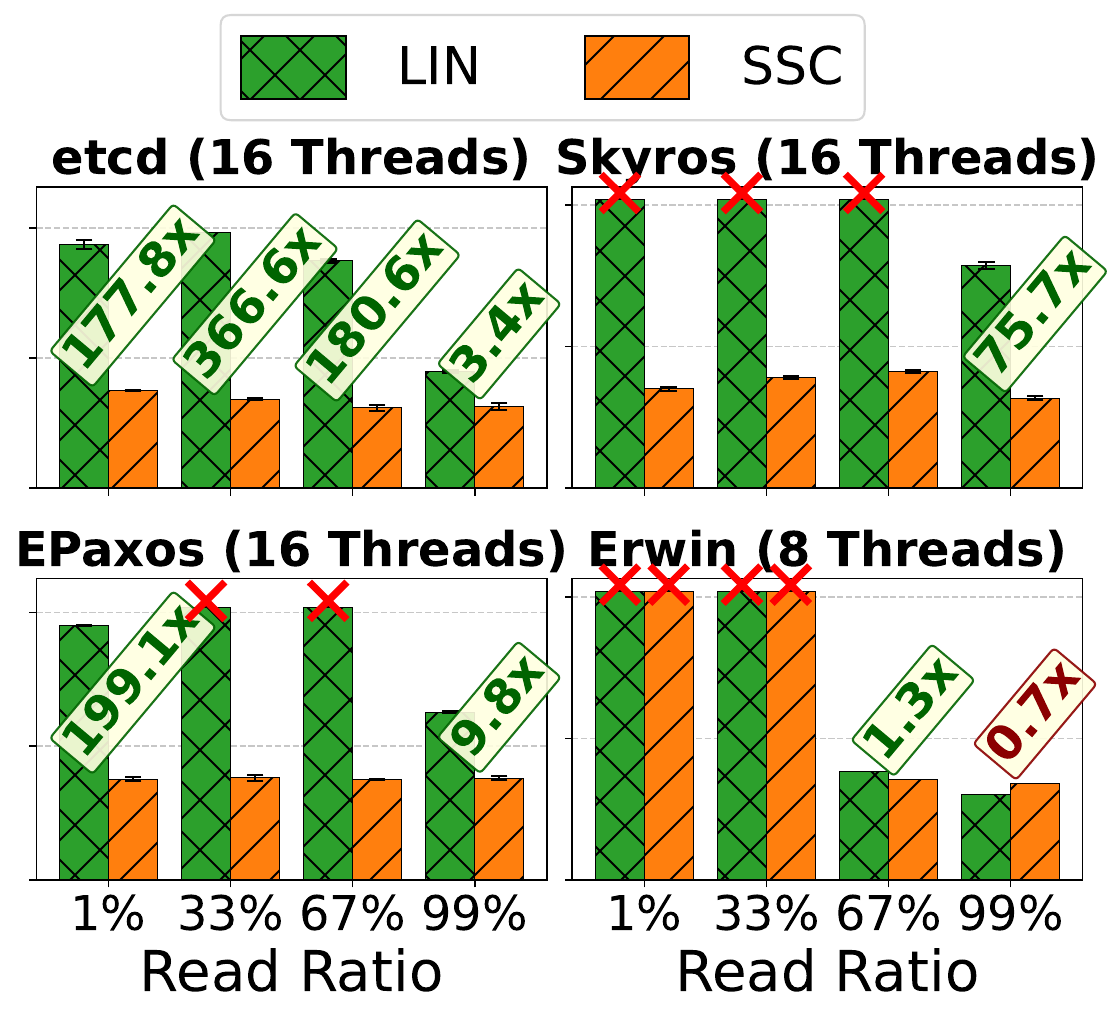}
    \vspace{-15pt}
\end{minipage}
\hfill
\begin{minipage}[t]{0.275\textwidth}
    \centering
    \includegraphics[width=\linewidth]{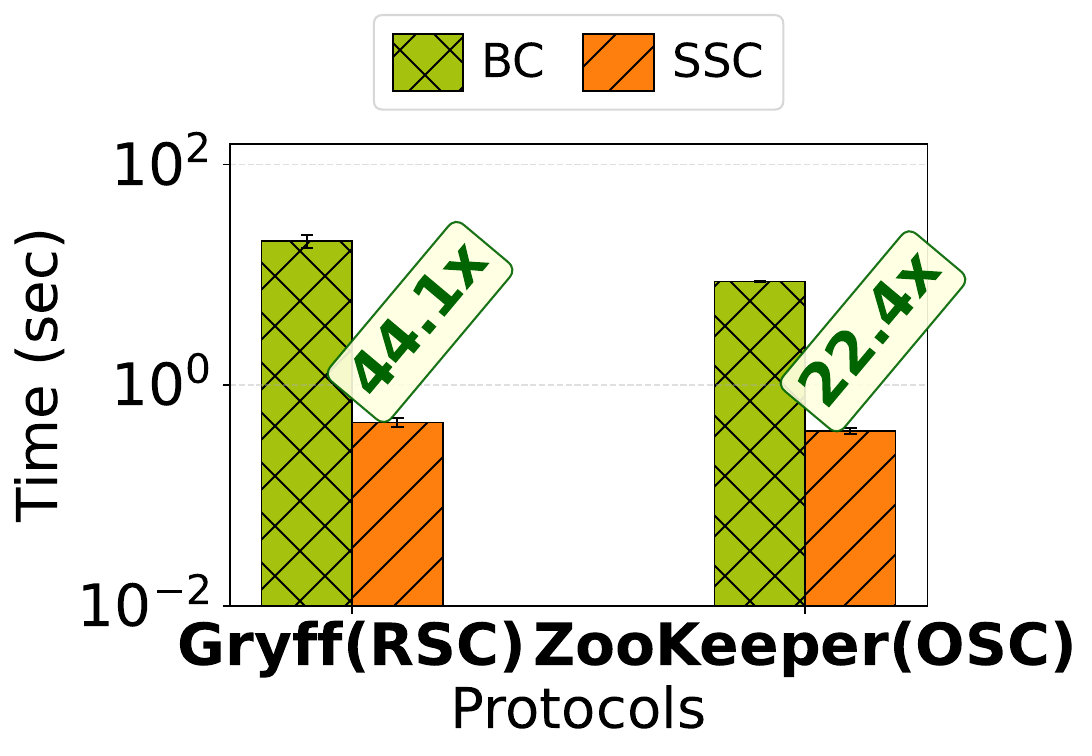}
    \vspace{-10pt}
\end{minipage}
\vspace{-15pt}
\caption{Performance comparison for checking basic consistency guarantees
(linearizability, ordered sequential consistency, and regular sequential
consistency) with checking for system-specific consistency guarantees.
System-specific consistency checking can be upto \faster faster and scale better
when increasing concurrency and reducing read ratios.}
    \label{fig:result_base}
\vspace{-10pt}
\end{figure*}

Figure~\ref{fig:result_base} compares the end-to-end runtime when checking for
linearizability, and when checking for system-specific consistency guarantee,
while varying the number of client threads and read ratios. Due to page limit, 
we only show results for two representative systems: etcd is a total order
system and Erwin is a lazy ordering system.

We observe that as
number of client threads increase, and when read ratios decrease, the end-to-end
runtime increases for checking linearizability. For higher concurrency and lower 
read ratios, linearizability checking starts timing out.

However, this is not always the case for system-specific consistency checking.
For instance, the time it takes to check for etcd-linearizability, as we
increase number of client threads and reduce read ratios, only grows from
0.14s to 1.87s. This is because etcd ordering oracle already provides a 
total order over all the operations using the revision number order hint. For
other systems, the time does grow (at a reduced rate) because their ordering
oracles only provide partial order constraints.  

This speed up is also observed when the basic consistency guarantee is not
linearizability, as shown in the rightmost plot in Figure~\ref{fig:result_base}.
In OSC and RSC, \name without system-specific ordering oracle quickly times out,
because these consistency guarantees are even weaker than linearizability, which
increases the search space of possible topological orders.

Using system-specific consistency guarantees, we found a consistency violation
bug in Gryff-RSC which we are still root-causing. For Zookeeper,
we did not find any consistency violation bug.  Note that
existing checkers, like Porcupine and Knossos, cannot check these system setups
as they do not guarantee linearizability.

{\em Takeaway: System-specific consistency checking can be upto \faster
faster when compared to the basic consistency checking. 
This makes system-specific consistency checking robust in evaluating
write-intensive and highly concurrent histories that traditionally bottleneck
existing checkers. \name generalizes to checking other consistency guarantees
like ordered sequential consistency and regular sequential consistency which
cannot be checked by existing checkers.}

\subsection{Impact of Ordering Oracle's Strength} 
\label{sec:eval-ablation}

We do two evaluations in this section. We compare the effectiveness of \name's
DAG-based algorithm with the etcd's modified history approach
(Figure~\ref{fig:etcd_adjusted}) for doing system-specific consistency checking.
Then, we evaluate the impact of soft order hints ($\soft$) on end-to-end
runtimes.

\begin{figure}[ht]
    \centering

    \pgfmathsetmacro{\total}{0.99}
    \pgfmathsetmacro{\algowidth}{0.52}
    \pgfmathsetmacro{\softwidth}{\total-\algowidth}

    \subfigure[]{
        \label{fig:result_algo_a}
        \includegraphics[width=0.51\linewidth]{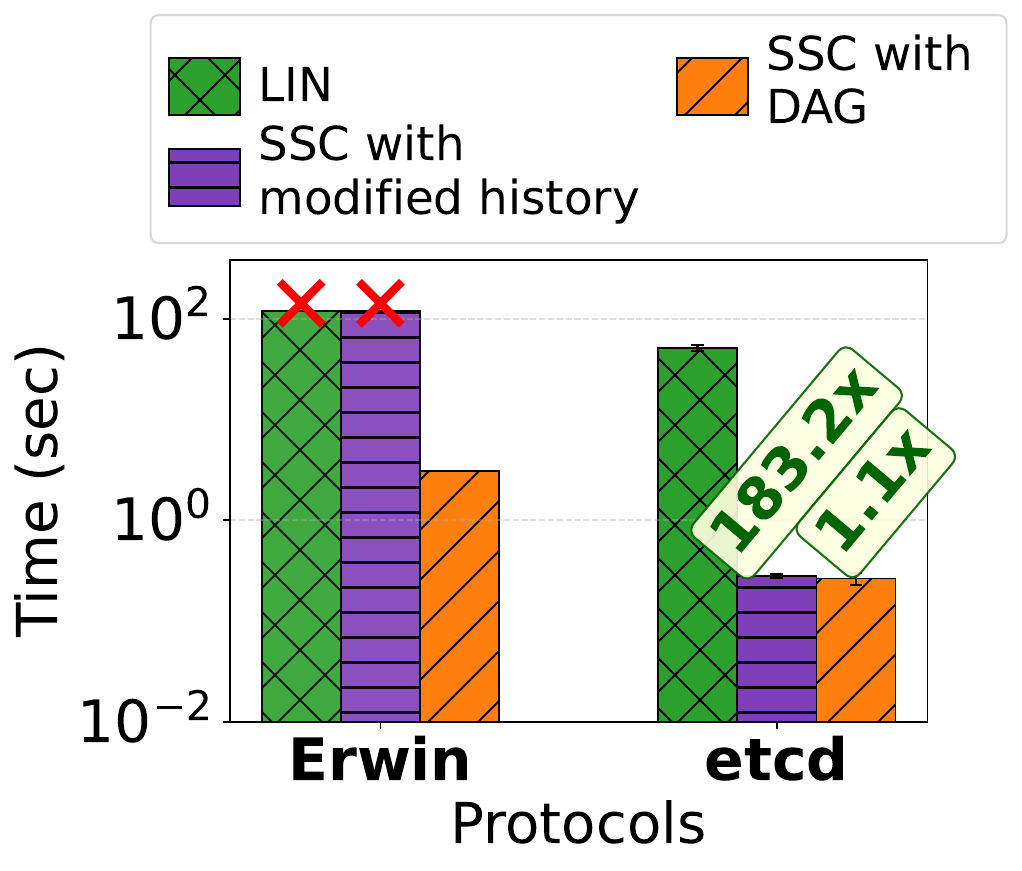}
    }%
    \subfigure[]{
        \label{fig:result_algo_b}
        \includegraphics[width=0.46\linewidth]{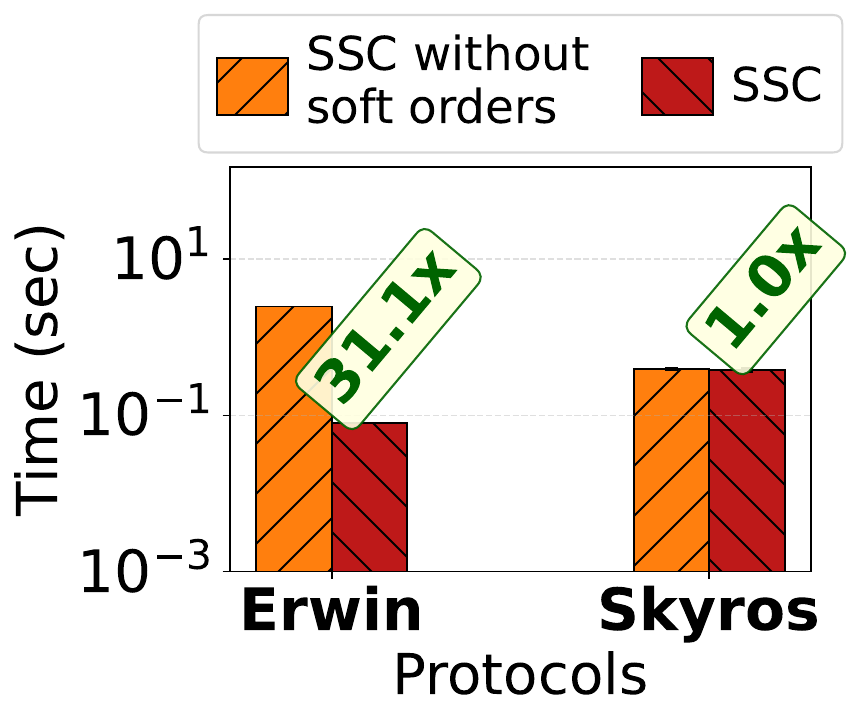}
    }

    \vspace{-10pt}
    \caption{(a) Modified history algorithm is effective for total-order hard
    constraints, such as in etcd, but not for Erwin. \name's DAG algorithm is
    effective for both systems as it does not lose ordering information.
    (b) Soft order constraints in system-specific consistency checking can
    provide modest speedups.}
    \vspace{-10pt}
    \label{fig:result_algo}
\end{figure}
Figure~\ref{fig:result_algo_a} compares the end-to-end runtimes for basic
consistency checking, and system-specific consistency checking using the
modified history approach and the \name's DAG-based algorithm. Due to page
limit, we again show results for the three representative systems.

We observe that the modified history approach already realizes the large speed
up over linearizability checking for etcd. This is because etcd's ordering
oracle provides total order constraints using revision numbers, which can be
fully captured by the modified history approach (recall
Figure~\ref{fig:etcd_adjusted}). However, \name's DAG-based algorithm gives
$>180\times$ speed up for Erwin and etcd, over the modified history approach.  This is
because the modified history approach loses direct ordering relations (\ie
non-transitive) because the modified history approach can only express interval
orders, whereas Erwin only provide partial order
constraints, which may not form an interval order. 

Figure~\ref{fig:result_algo_b} removes just the soft orders from Erwin ($\soft$ in
Table~\ref{tab:protocol_impl}), to measure the impact of soft order on
end-to-end runtimes. We observe that soft orders improve runtimes by $1.2\times$
in Erwin.

These time differences can be understood by the number of stack pops that happen
while serializing the history, as is also shown in Figure~\ref{fig:result_algo_b}. While hard ordering constraints ($\hard$) reduce the number of possible
topological orders, soft ordering constraints ($\soft$) help \name prioritize
among concurrent operations.

{\em Takeaway: Hard order constraints ($\hard$) reduce the search space and soft
order constraints ($\soft$) help prioritize the search. \name's DAG algorithm 
does not lose ordering relations, which are lost by the modified history
algorithm. This leads to $183\times$ speed up over the modified history
algorithm.}

\subsection{Overhead}
\label{sec:overhead}

Finally, we compare the execution time and peak memory consumption of \porcupine
against \name for linearizability checking \ie \textit{without} utilizing any
order hints. 
We report results for operation histories with concurrency set to 16 \& 10
respectively for etcd and Erwin and with read ratio 0.5.

\begin{figure}[ht]
    \centering
    
    \includegraphics[width=0.55\linewidth]{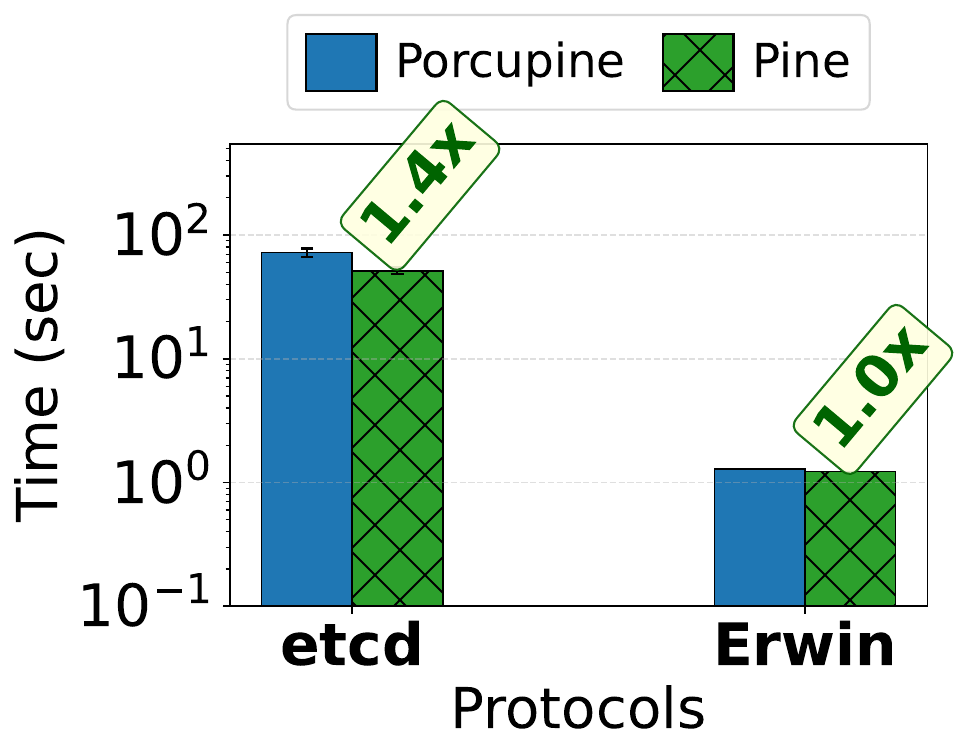}
    \vspace{-10pt}
    \caption{\name provides comparable performance to \porcupine for testing
    linearizability.}
    \label{fig:result_overhead}
    \vspace{-10pt}
\end{figure}

As shown in Figure~\ref{fig:result_overhead}, when checking linearizability (\ie
no order hints), \name matches the execution speed of \porcupine.  This confirms
that \name's preprocessing steps, discussed in Section~\ref{sec:hintalgo}, adds
negligible runtime overhead.  

We found that \name's memory consumption was always lower than that of
\porcupine, both when \name was checking linearizability and when it was
checking system-specific consistency guarantees. This is because for both the
implementations, the memory consumption is not dominated by the operation
history, but rather a bitset that caches the linearizations tried thus far.

{\em Takeaway: \name does not add much time or memory overhead over
\porcupine, while providing more general consistency checking algorithm that
goes beyond linearizability and supports system-specific consistency guarantees.}

\section{Related Work}
\label{sec:related}

Storage system errors can lead to severe consequences, such as data
loss~\cite{simple_testing_osdi14}, service
unavailability~\cite{cosmos_tla_icse23, recovery_hotos_2013, aws_tinydb_nsdi20}, loss of user
trust and revenue~\cite{failslow_fast18, aws_outage17}.  These errors can be due to
both protocol bugs~\cite{cosmos_tla_icse23, epaxos_tla} and implementation
bugs~\cite{explode_osdi10,simple_testing_osdi14}. Protocol bugs have been
targeted by formally verifying safety and liveness properties of the system's
specification~\cite{crash_hoare_sosp15, microsoft_ccf, mongo_tla}.
Implementation bugs have been targeted by validating the conformance of the
implementation with its
specification~\cite{aws_lightweight_sosp21} using proof
checkers~\cite{ironfleet}, extracting implementation from the
specification~\cite{verdi_pldi15}, and using model checkers~\cite{explode_osdi10}, refinement
checkers~\cite{rprc_nsdi25}, and trace validation~\cite{ccf_trace_validation,
etcd_trace_validation} while performing property-based testing~\cite{chess},
fuzzing~\cite{randomized_testing_2018} and fault-injection testing~\cite{jepsen, aws_lightweight_sosp21} over the implementation.

Consistency checkers remain prevalent for checking distributed storage
systems~\cite{memorydb_sigmod24, etcd_uses_porcupine, legostore_vldb22,
tidb_vldb20} as they directly check what is observed by the clients. Consistency
checkers do not necessitate writing a proof for the
implementation~\cite{ironfleet}, limiting the implementation
language~\cite{verdi_pldi15}, writing a complete specification~\cite{rprc_nsdi25,
mongo_tla}, developing new logics~\cite{crash_hoare_sosp15}, keeping the
implementation and specification always in-sync with one
another~\cite{aws_lightweight_sosp21}, and making assumptions about the
correctness of unverified libraries, unsafe code, and possible system
errors~\cite{empirical_study_eurosys17}. \name performs system-specific
consistency checking which requires lightweight annotation with order hints on
responses: $<100$ lines in all the studied systems.

We have already discussed linearizability checkers in Section~\ref{sec:back}.
There are also transactional consistency checkers like Elle~\cite{elle}, which are 
orthogonal to this work on checking non-transactional consistency guarantees.
The non-transactional consistency checkers, built at
Facebook~\cite{existential_cons} and HP Labs~\cite{pahoehoe}, have differing
goals than \name.  These checkers count how often ``consistency anomalies" are
observed in their eventually consistent storage systems, whereas \name,
Porcupine, and Knossos are typically useful to catch implementation bugs.

Both the Facebook and HP Labs checkers extend beyond checking linearizability,
but they only extend to ``local'' consistency guarantees (\eg per-object
sequential consistency, and read-after-write consistency for the Facebook's
checker), where the input operation history can be partitioned into smaller
histories (\eg operations to each key in a key-value store), each of which can
then be checked independently.  However, operations on many data types, \eg a
totally ordered shared log~\cite{lazylog}, cannot be partitioned into
independent histories.  Further, several consistency guarantees, like
ordered sequential consistency, are non-local. Such consistency
guarantees can not be checked with their checkers.

Their checking algorithms only works for shared registers with \code{put} and
\code{get} operations, as they match \code{get} operations with their corresponding 
\code{put} operations.
Therefore, the storage systems that provide read-modify-write
operations~\cite{gryff, sosp21_rsc, redis}, \eg increment a counter, cannot be checked
with their checker.


\section{Conclusion}
\label{sec:conc}
In this paper, we presented a consistency checking approach that generalizes
beyond just checking for linearizability. This was possible because of the
decomposition of linearizability into more primitive \code{SingleOrder},
\code{RVal}, and \code{RealTime} properties by
Burckhardt~\cite{principles_of_ec}. Our algorithm generalizes \code{RVal} into
validity properties and \code{RealTime} into ordering properties. With this 
generalization, we could check for weaker consistency guarantees like ordered
sequential consistency and regular sequential consistency.

We could also strengthen consistency checking towards system-specific
consistency guarantees, where systems provide system-specific hard and soft
ordering constraints, over and above their basic consistency guarantee. Our
evaluations show that checking for system-specific consistency can be upto 
\faster faster, and can scale better to challenging scenarios of higher
concurrency and lower read ratios. We report \bugs new bugs, out of which
\uniqbugs could not be found with existing consistency checkers.

\bibliographystyle{plain}
\bibliography{eurosys27}


\end{document}